# Tough and high-temperature stable nacre-like $Bi_4Ti_3O_{12}$-based piezoceramics

Ruxue Yang[1], Temesgen Tadeyos Zate[2], Elo Overgaard Mogensen[2], Astri Bjørnetun Haugen[2], Florian Bouville[1]

1 Centre for Advanced Structural Ceramics, Imperial College London, London, United Kingdom

2 Department of Energy Conversion and Storage, Technical University of Denmark, Agnes Nielsens vej, Building 301, 2800 Kgs Lyngby, Denmark

*Abstract*

$Bi_4Ti_3O_{12}$-based ceramics are promising candidates for high-temperature piezoelectric devices owing to their high Curie temperature (> 600 °C). However, their low piezoelectric constants and poor mechanical reliability hinder their use in some engineering applications. Here, we fabricate nacre-like $<00l>$-textured ($f_{00l}$ > 94%) $Bi_{3.96}Ce_{0.04}Ti_{2.965}W_{0.0175}Nb_{0.0175}O_{12}$ (NL-BCWNT) ceramics via a scalable magnetic-assisted slip casting (MASC) self-assembly process. In addition to crystallographic texture, the final material presents a brick-and-mortar microstructure with micro-sized grains with a median aspect ratio of 15. NL-BCWNT exhibits a high fracture toughness for piezoceramics, with both $K_{IC}$ and $K_J$ reaching 2.2 ± 0.4 MPa·m$^{0.5}$ and 4.2 ± 0.4 MPa·m$^{0.5}$, respectively. The toughening originates from deflection and stable crack propagation within the nacre-like structure. The $d_{33}$ is 30 ± 3 pC/N with a Curie temperature of 661°C, and $d_{33}^*$ reaches 46 ± 3 pm/V at 160 °C with less remanent strain than random BCWNT. The improvements are attributed to aligned domain along spontaneous polarization direction $<h00>$ with better mobility. NL-BCWNT exhibited excellent ferroelectricity fatigue resistance with higher than $10^6$ cycles without degradation. Furthermore, similar improvements are observed in pure BiT, suggesting the general applicability of this strategy to other BiT-based systems. The nacre-like architecture thus provides a promising addition to the design toolbox for high-performance electromechanical materials.

*Introduction*

High-temperature stable piezoelectric ceramics are critical for devices operating in extreme environments, including sensors for non-destructive evaluation in energy generator (nuclear, geothermal), high-precision speed sensing in aerospace or actuators for fuel injectors for automotives.[1,2] These demands require piezoelectric materials with high Curie point $T_C$, large piezoelectric coefficients (e.g. $d_{33}$) and adequate mechanical performance.[3] Currently, $Pb(Zr_xTi_{1-x})O_3$-based ceramics dominate the market due to their high $d_{33}$ (>400 pC/N).[4] However, their relatively low $T_C$ (< 350 °C) limits operation in harsh environments[4], and their moderate fracture toughness ($K_{IC}$ < 2 MPa·$m^{0.5}$)[5] limit their lifetime. In addition, the lead content raises environmental and safety concerns.[6] Therefore, the development of mechanically stronger lead-free piezoelectric materials capable of maintaining piezoelectric performance at high temperatures remains urgently needed.

Bismuth layer-structured ferroelectrics (BLSFs), especially Aurivillius-type $Bi_4Ti_3O_{12}$-based (BiT-based) ceramics, are promising candidates owing to their high $T_C$ (~675 °C), large spontaneous polarization, excellent thermal stability, and intrinsically low dielectric permittivity.[7] However, its $d_{33}$ remains limited, reaching only 6-8 pC/N.[7] Various strategies have therefore been proposed to enhance the piezoelectric performance of BiT. Oxygen vacancies induced by $Bi^{3+}$ volatilization can lead to high conductivity and difficult poling.[8] Donor doping, such as $W^{6+}$/$Nb^{5+}$/$Ta^{5+}$/$Sb^{3+}$ on A/B sites, mitigates these defects by suppressing vacancies, increase resistivity and promote domain switching, increasing $d_{33}$ up to 20~30 pC/N.[9] Within the highly anisotropic layered Aurivillius phase, $< h00 >$ is the spontaneous polarization direction, and increased domain wall mobility is observed when the electric field is applied along this direction.[10] Producing crystallographically textured samples allows to take advantage of the crystal anisotropy at the sample scale, with $< 00l >$-textured BiT reaching 14 pC/N perpendicular to the texture direction.[10] Domain configuration tailoring has also emerged as an effective strategy to enhance piezoelectric performance. In Ce and W/Nb co-doped $Bi_{4-x}Ce_xTi_{3-2y}W_yNb_yO_{12}$ (BCWNT) ceramics, reduced domain size and increased number of 180° domain walls improve the overall domain-wall mobility during poling, leading to an increased $d_{33}$ of 38.5 ~ 40.2 pC/N.[7,11]

However, all the above reported achievements did not change the BiT-based material's mechanical performance. Even worse, increased dopant content leads to weaker mechanical performance, with the BCWNT composition showing a flexural strength of only of 29 ± 2 MPa and fracture toughness of ~0.59 MPa·$m^{0.5}$, 68% and 63% smaller than undoped BiT, respectively.[12] Overcoming the mechanical limitations of BiT-based ceramics would encourage their practical uses. However, conventional toughening strategies, such as incorporating second-phase particles, whiskers, or microstructural modification typically provide only limited mechanical improvement and often compromise piezoelectric / ferroelectric performance.[3] Considering the inherently low piezoelectric response of BiT, such trade-offs are difficult to justify. Systematic mechanical reinforcement strategies for piezoelectric ceramics remain scarce.

In devices, the direction for piezoelectric signal output or input might be different from the direction in which cracks will grow and cause component breakdown. In bulk materials, the commonly used piezoelectric coefficient $d_{33}$ corresponds to the applied mechanical load (or electric field) parallel to the induced electric displacement (or strain). Simulations and experimental results indicate that, under such loading conditions, bulk ceramics generally exhibit higher probability for microcracks growing with deflection angle several tens of degrees away from the loading direction, because of the mechanical stress concentration, charge accumulations or domain switching induced strain.[13,14] In some devices, such as vibration cantilever-beam transducer, the maximum piezoelectric signal output occurs near the free end during bending, whereas the highest mechanical stress concentrations occur at the outermost tensile and compressive surfaces.[15] These regions are also one of the origins of mechanical degradation in cantilever beam transducers.[16] Therefore, for some piezoelectric applications, it might be beneficial to decouple functional and mechanical orientation, enabling enhanced piezoelectric response along one axis and improved mechanical properties another.

Owing to the layered Aurivillius structure, BiT can be synthesized as high aspect-ratio platelets via molten-salt methods.[17] This feature provides an ideal platform for constructing nacre-like microstructure, i.e. microstructure with platelet-shaped particles with their plate normal aligned in a common orientation during processing and sintering. In structural ceramics such as $Al_2O_3$[18], nacre-like architectures have demonstrated simultaneous strengthening and toughening, primarily through crack deflection along the direction normal to the aligned platelet plane[18]. These nacre-like architecture can be fabricated via multiple processes. Among all these processes, magnetically assisted slip casting (MASC)[18] is a scalable and versatile process that can form complex shaped bulk part in which micron-sized nano-$Fe_3O_4$-coated anisotropic building blocks are assembled with their orientation controlled using a low intensity magnetic field[18].

Here, we fabricate a highly $<00l>$-textured $Bi_{3.96}Ce_{0.04}Ti_{2.965}W_{0.0175}Nb_{0.0175}O_{12}$ (BCWNT) ceramic with a nacre-like architecture (NL-BCWNT) via MASC. The BCWNT platelets produced by molten salt synthesis have their crystallographic $<00l>$ axis ($c$-axis) aligned along the platelets thickness. MASC allow us to form bulk sample with a high-quality fibre texture (Lotgering factor $f_{00l}$ =94%) by aligning the platelets along a chosen direction. This architecture enables a simultaneous enhancement of mechanical and piezoelectric performance. We demonstrate that these textured samples present stable crack growth leading to improved fracture toughness and strength when the crack is introduced in the direction of the texture. Meanwhile, the texture amplifies domain switching and consequently increases the ferroelectric and piezoelectric response in the direction perpendicular to the texture. The increased toughness restrains microcrack development, reducing mechanical degradation and resulting in improved long-term ferroelectric cycling performance. Importantly, the proposed nacre-like design relies on architectural design rather than compositional modulation and can thus be applied to other compositions that can be synthesised as micro-sized anisotropic particles.

*Main text*

We used magnetically assisted slip casting (MASC) with BCWNT platelets to form $< 00l >$ $-$fibre textured sample with nacre-like microstructure (NL-BCWNT). BCWNT platelets from molten salt synthesis (Figure 1a) present a median aspect ratio of 15 and median thickness of 0.22 μm (Table S1). The thickness direction of these platelets corresponds to the $< 00l >$-axis. The presence of dopant in the platelets is qualitatively confirmed using EDX (Figure S1). After the synthesis, the platelets were coated with < 0.1 wt% of $Fe_3O_4$ to increase their magnetic susceptibility[19]. As shown in Figure 1b, we then dispersed nano-$Fe_3O_4$-coated BCWNT platelets in ethanol to form a slurry. This slurry was cast in a mould put on top of a gypsum base in front of a rotating 500mT magnet. The platelets aligned in the plane formed by the rotating magnetic field. These horizontally aligned platelets deposited on the gypsum substrate and then dried into green bodies (Figure 1c).[18] The green bodies were densified by Sparkling Plasma Sintering (SPS) reaching relative density of 96.6% ± 0.6%. This ethanol-based MASC process is safe, non-toxic and suitable for fabricating bulk parts.

We used SEM and electron backscatter diffraction (EBSD) to quantify the nacre microstructure and crystallographic texture of the sintered NL-BCWNT samples. Inverse pole figure (IPF) map along the texture direction on SEM image of the cross section perpendicular to the texture direction displays highly aligned anisotropic grains with similar crystallographic orientation (Figure 1d). The pole figure shows strong $< 00l >$ texture, further confirming the $< 00l >$ fibre texture (Figure 1e). The pole figure shows that the clustered intensity has a full width at half maximum of 29-33°, indicating that the platelets in the NL-BCWNT ceramics are not perfectly aligned but exhibit a finite angular spread around the $< 00l >$ direction. This texture can also be seen in the bulk sample as shown in the X-ray diffractograms (XRD) of the NL-BCWNT and reference un-doped BiT (Figure 1f). NL-BWCNT exhibits a higher relative intensity of the $[00l]$ family diffraction peaks compared with the reference BCWNT, leading to a Lotgering Factor higher than 94%.

To summarize, we successfully produced a dense, $< 00l >$-fibre textured NL-BCWNT with a nacre-like microstructure ceramic and we can now measure the effect of this microstructure on the sample's mechanical and piezoelectric properties.

As specimens were tested in multiple orientations, we define the testing configurations as follows to ensure clear and consistent notation:

***Mechanical:*** To characterise the transverse isotropic mechanical behaviour in bending of NL-BCWNT, specimens testing direction are denoted using two letters, where the first letter indicates the tensile stress plane normal direction or crack plane normal direction, and the second the direction of maximum opening stress or the crack propagation direction in four-points bending and SENB tests respectively. Here, $\boldsymbol{l}$ denotes the longitudinal direction corresponding to the texture direction, and $\boldsymbol{t}$ denotes the transverse direction (any direction perpendicular to the texture direction). Mechanical tests were performed on BCWNT-$tl$ and BCWNT-$tt$, revealing pronounced mechanical anisotropy. Fracture strength was measured with four-point-bending tests. Fracture toughness was measured using single-edge notched beam (SENB) tests respectively, where the displacements and crack length have been measured optically to ensure their accuracy.

*Piezoelectric and Dielectric:* The two testing configurations are denoted as BCWNT-$l$ and BCWNT-$t$, respectively (Figure 3a). The "-$l$" refers to the configuration in which the electric field and the applied force are both along the longitudinal direction, parallel to the texture direction. The "-$t$" refers to the configuration in which electric field and the applied force are transverse to the texture direction.

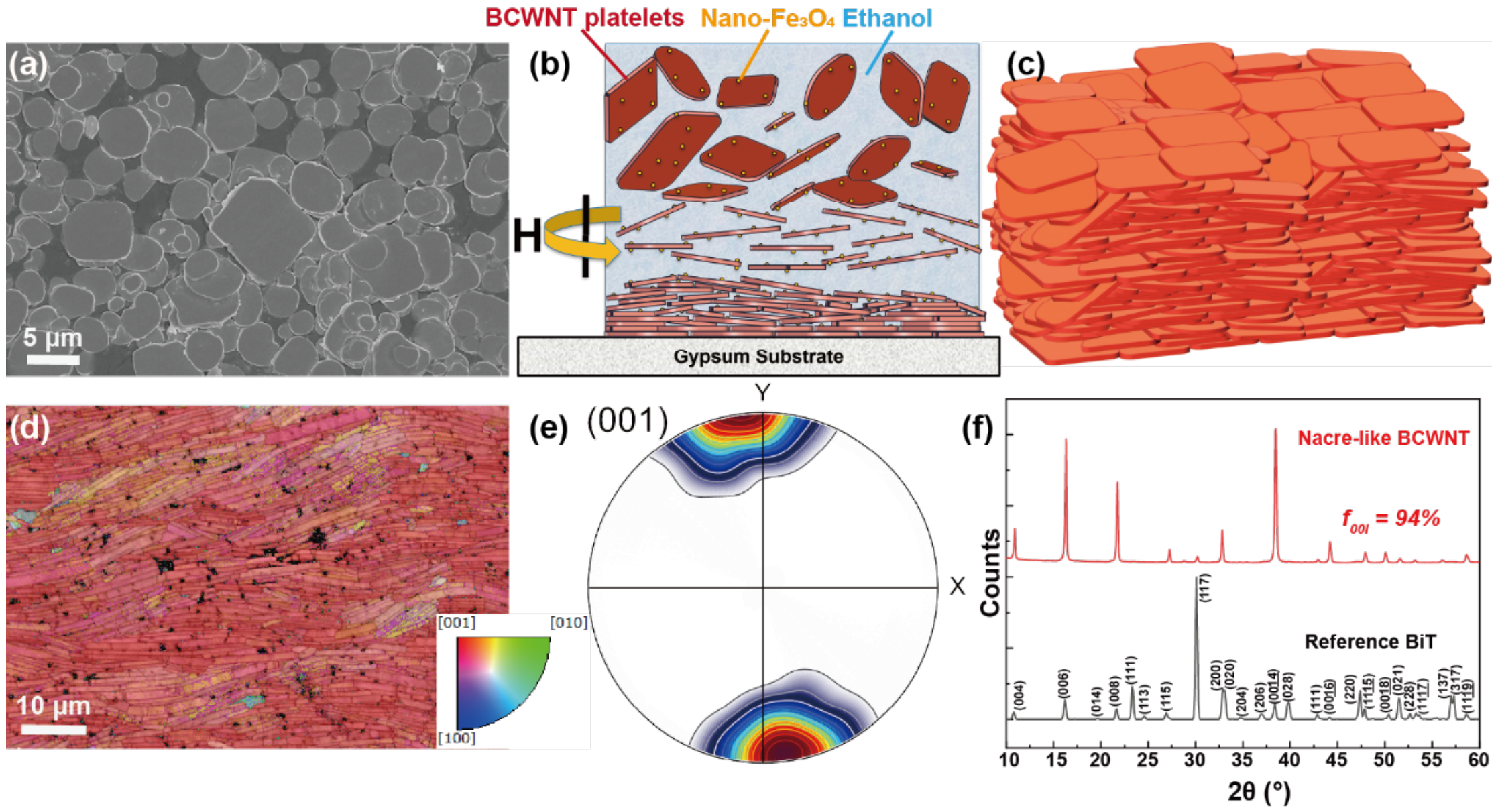

**Figure 1. Nacre-like BCWNT (NL-BCWNT) processing and microstructure.** (a) Scanning electron microscopy (SEM) images of BCWNT platelets powder prepared by molten salt synthesis. (b) A schematic of magnetically assisted slip casting (MASC) with nano-$Fe_3O_4$-coated BCWNT platelets. (c) Illustration of green body structure after MASC. (d) SEM Electron backscatter diffraction (EBSD) maps (inverse pole figure along the Y direction) taken on a cross section perpendicular to the texture direction and (e) corresponding pole figure pattern of $(00l)$ plane. (f) XRD pattern for reference BiT and NL-BCWNT.

As shown in Figure S2, all samples exhibit stress-strain curves with linear elastic brittle behavior. In Figure 2a, these tests reveal that NL-BCWNT strength is 80 $\pm$ 3 MPa for BCWNT-$tl$ and 90 $\pm$ 20 MPa for BCWNT-$tt$, corresponding to improvements of 267% and 300% compared with the same composition ceramic (29.66 MPa) in previous reports[12]. Due to the lack of direct experimental evidence, we do not exclude the possibility that ferroelasticity may affect the mechanical model used for strength calculation. However, a detailed quantitative evaluation of ferroelastic effects is beyond the primary scope of this study and is not present in the literature. Since the experimental data still shows a clear linear trend in Figure S2, we have tentatively adopted the calculation model for normal ceramic materials to estimate the strength. This improvement in strength may be attributed to two factors. First, SPS sintering can produce a sample with smaller defects. Second, nacre-like architecture may delay crack initiation. Like natural nacre [20–22] and NL-$Al_2O_3$[23], the fracture modes changes from conventional Mode I to mixed mode due to the deflection of the crack on the microstructure. Such fracture mode can require higher applied stress intensities to initiate a crack.[24]

During SENB tests, BCNWT-$tl$ shows stable crack growth with R-Curve behaviour (Figure 2b). The stress intensity factor $K_J$ increase with crack extension up to 0.5 mm, consistent with the non-linear behaviour obtained in the force-displacement curves (Figure S3). BCWNT-$tl$ sample present a toughness at crack initiation of $K_{IC}$ = 2.2 ± 0.4 MPa×m$^{0.5}$ and a final $K_J$ up to 4.2 ± 0.4 MPa·m$^{0.5}$ after a crack extension of 0.5 mm (Figure 2c), a 90% increase from $K_{IC}$ and a 2.6 to 7-folds increase compared to the reported random BCWNT ceramics (0.6 ~ 1.6 MPa×m$^{0.5}$)[12]. The BCWNT-$tl$ fracture surfaces shows (Figure 2d): (1) at the sample scale, the crack path deflects with an average angle 83° ± 6° from the initial notch for about 1.7 mm, followed by a tortuous path along the maximum opening stress direction (Figure 2d); (2) at the grain scale, the fracture surface reveals a staircase morphology (Figure 2d). Large, flat grain facets are exposed, accompanied by pull out, indicating the presence of crack deflection, bridging and pull-out toughening mechanisms. Transgranular fracture also occurred as we can observe exposed platelets cross section on the fracture surface.

These observations suggest that: (1) crack initiation is delayed to higher stress intensity as the orientated grains prevent the crack from propagating directly in mode $I$, instead forcing the crack along the grain boundary at a deflected angle of 83° ± 6° and in mixed-mode.[25,26] (2) As the crack propagates, it experiences repeated deflections along the well-aligned grain boundaries and grains. The enhanced crack deflection and bridging mechanisms associated with the nacre-like architecture triggers stable crack growth over an extended crack extension, thereby leading to increased $K_J$ and a rising R-curve response.[27,28]

Non-brittle fracture behaviour and R-curve characteristics have also been reported in several piezoelectric ceramics, such as PZT- and BZT-based ceramics.[3] However, the toughening mechanism in these systems mainly relies on ferroelastic toughening within the grains. To enable a direct comparison of the effectiveness of these toughening mechanisms, we calculate the toughness amplification as $\Delta K = K_J - K_{IC}$.[29] Previous studies have shown that the $\Delta K$ from ferroelastic toughening is usually around 0.1 ~ 0.5 MPa×m$^{0.5}$, [3] whereas the NL-architecture can provide a 4 times higher toughness amplification, reaching up to ~2 MPa×m$^{0.5}$. This suggests that the NL-architecture is more effective in enhancing fracture toughness than previous mechanisms used in piezoceramics.

Finally, despite adopting a similar nacre-like structure, NL-BCWNT exhibits a different fracture behavior from that of the nacre-like $Bi_{0.5}Na_{0.5}TiO_3$ (NL-BNT) ceramics with a $SiO_2$ as intergranular mortar phase reported in our previous studies[30]. For NL-BNT, we achieved a toughness improvement from 0.9 MPa×m$^{0.5}$ to 1.8 MPa×m$^{0.5}$ without sacrificing piezoelectric performance.[30] However, the material remained brittle, and the crack was found to propagate directly through the grains[30]. We assume that this difference between NL-BCWNT and NL-BNT may be attributed to two possible factors: (a) BNT single crystal in the NL-BNT is relatively weak[30], and limited contrast can be obtained between the grain and the $SiO_2$ mortar ($K_{IC} \approx$ 0.7 MPa×m$^{0.5}$)[31]. In contrast, NL-BCWNT may possess a sufficiently large toughness contrast between the grains and grain boundaries, deflecting the crack propagation along the grain boundaries. This contrast can originate from the anisotropic lattice structure of BCWNT compared with the more isotropic one of BNT. In monocrystal, fracture toughness can be linked with surface energy using Griffith relationship[32] $K_{IC} = \sqrt{2\gamma E}$, where $\gamma$ is the surface energy of the new surface created during the fracture and $E$ is the Young's modulus. While it has not been measured, the $(00l)$ planes in BCWNT are supposed to present a lower surface energy than the $(hk0)$ as suggested by the high aspect ratio particles presenting mainly the $(00l)$ planes produced by molten salt[17,33]. This is similar to nacre-like alumina, where the surface energy of the $(00l)$ surface is low compared to the $(hk0)$[34,35] and cracks also deflects parallel to the $(00l)$. By contrast, BNT adopts a rhombohedral $R3c$ structure at room temperature that forms from a cubic $Pm\bar{3}m$ perovskite structure, in which the lattice and thus the surface energy along the parent crystallographic axes are relatively similar. Consequently, its crystallographic anisotropy is comparatively weak, and the tendency for direction-dependent crack propagation is suppressed. However, validating these hypotheses would require dedicated experiments to measure or simulate BCWNT crystallographic plane surface energies. A rigorous investigation of this issue would constitute a research topic beyond the scope of the present work.

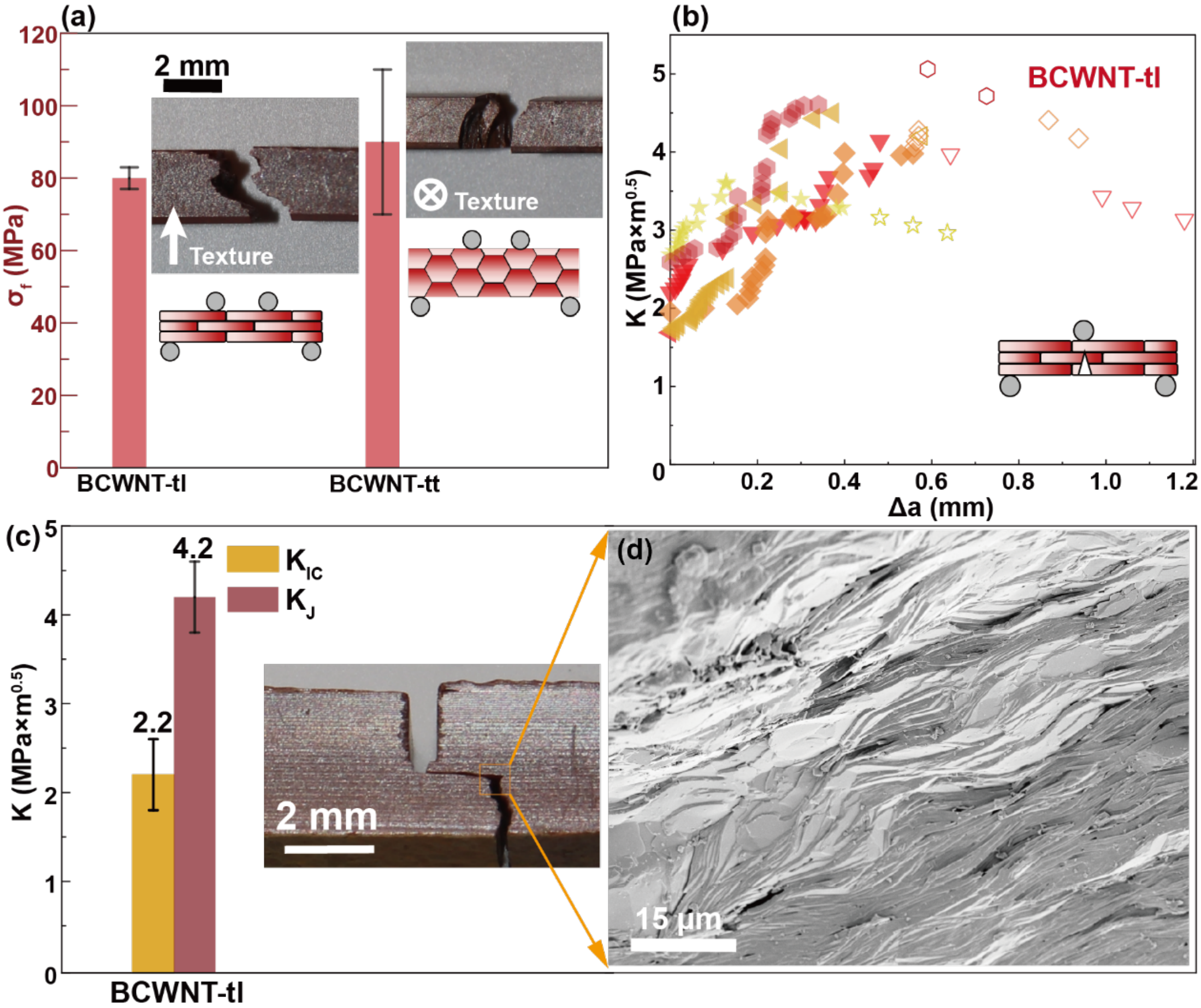

**Figure 2 Mechanical performance improvement from nacre-like microstructure.** (a) Four-point-bending test results for NL-BCWNT in the $tl$ () and $tt$ direction. Inset shows representative fracture of the samples in the two directions (b) R-curve of BCWNT-$tl$. Hollowed symbols represent crack propagation values that fall outside the ASTM Standard recommended limit. (c) Critical stress intensity factor at initiation $K_{IC}$ and after crack propagation to the ASTM standard limits $K_J$ for BCWNT-$tl$. (d) SEM image of the fracture surface, inset indicates where the image was taken and the crack path on a representative sample.

Following our discussion of the mechanical performance improvements of the textured BCWNT ceramics, we further analyze their ferroelectric and piezoelectric properties. The goal is to measure the effect of texture and to study the link between mechanical and functional enhancements.

Figure 3a compares the temperature-dependent dielectric permittivity measured at 1 MHz of the two different directions (BCWNT-$l$ / $t$) within the nacre-like architecture. There are distinct and sharp permittivity peaks observed in both directions, indicating the Curie point $T_C$. The peak occurs at 661 °C for both BCWNT-$l$ and -$t$ and marks the transition from the ferroelectric orthorhombic phase to the paraelectric tetragonal phase. The relative permittivity values for BCWNT-$t$ are the highest across the entire temperature range, consistent with the other textured BiT-based systems[10,36]. A similar trend is also observed for the dielectric loss (Figure S4). This anisotropy can be attributed to the intrinsic resistivity anisotropy of the Aurivillius phase. The fluorite layers $[Bi_2O_2]^{2+}$ have higher resistivity, limiting charge interlayer transports along $< 00l >$.[37] In contrast, the pseudo-perovskite layers parallel to $ab$-plane support mixed ionic-electronic conduction at elevated temperature, giving a larger charges response.[38,39] The spontaneous polarization along the $< h00 >$ directions also enhance the transverse dielectric polar responses.

In Figure 3b, at room temperature, BCWNT-$t$ exhibits a $d_{33}$ value of 30 ± 2 pC/N. Under an identical applied field of 8 kV/mm, it is 119% higher compared to BCWNT-$l$ (13.7 ± 0.5 pC/N), and almost 5-folds of that of undoped BiT (~6 pC/N)[40]. The BCWNT-$t$ generates a stronger piezoelectric response compared with BCWNT-$l$ which is in accordance with previous studies on BiT-based structures where increased piezoelectric response is found parallel to the spontaneous polarization direction $< h00 >$.[10,36] Furthermore, based on previous research[7,11], this multi-element doping can (1) increase lattice distortion of the $TiO_6$ octahedra and asymmetry of the lattice parameters, enhancing the intrinsic contribution for the poling and piezo response;[7] (2) introduce higher density of 180° domain walls parallel to the $< h00 >$ direction, separating the domains into smaller ones with faster movement, enhancing the extrinsic contribution for more completed domain switching.[7] Furthermore, in the *ex situ* temperature-dependent $d_{33}$ testing (Figure 3c), BCWNT-$t$ showed some depolarization with increasing annealing temperature and a sharp drop at ~650 °C, consistent with the observed $T_C$ at ~661 °C.

Upon heating, the P-E loops of BCWNT−$\boldsymbol{t}$ become progressively developed, with $E_c$ increasing from 3.1 kV/mm (40 °C) to 4.5 kV/mm (160 °C) and $P_r$ and $P_{max}$ reaching 20 and 25 μC/cm², respectively (Figure 3d). The simultaneous increase in $E_c$ and the bulging at the loop tips indicate that the switching process remains strongly influenced by temperature dependent defect or conductivity effects [41]. For temperature-dependent unipolar S-E loops (Figure 3e), no obvious remnant strain is observed for BCWNT-$t$ up to 160 °C. The linearity and recoverable strain are advantageous for high-temperature actuator applications. Calculating $d_{33}^*$ from these loops showed gradually increased response in BCWNT-t (Figure 3f) from 34 ± 4 pm/V at 40 °C up to 46 ± 3 pm/V at 160 °C.

The ferroelectric bipolar P-E fatigue tests were also conducted for BCWNT-$t$ up to $10^6$ cycles at 180 °C. BCWNT-$t$ exhibits no signs of fatigue degradation; instead, its $P_r$ continues to increase throughout cycling, from 31 μC/cm$^2$ to approximately 34 μC/cm$^2$, a 10% increase, after $10^6$ cycles. The slight increase in $P_r$ at the early cycling stage has also been reported in various ferroelectric systems.[42–44] This behavior is commonly attributed to the de-aging of defects: oxygen vacancies and associated charged dipoles rearranging during the first cycles, and the external applied fields might drive these defects depinning and redistribution, thereby restoring local domain-wall mobility and enhancing the polarization response. Normally, this transient enhancement would be eventually overtaken by the dominant longer-term fatigue behavior, resulting in a decline in polarization.[42–44] The superior resistance to fatigue of BCWNT-$t$ can be attributed to two factors: (1) The texturing aligns the $ab$-plane with the electric field, enabling easier rotation of the $(h00)$-plane-parallel 180° domain walls and reducing the residual strain generated during non-180° domain-wall reorientation. (2) The enhanced toughness along $< 00l >$ suppresses preferential crack initiation deviated to the field and propagating during cycling, thereby mitigating mechanical degradation.

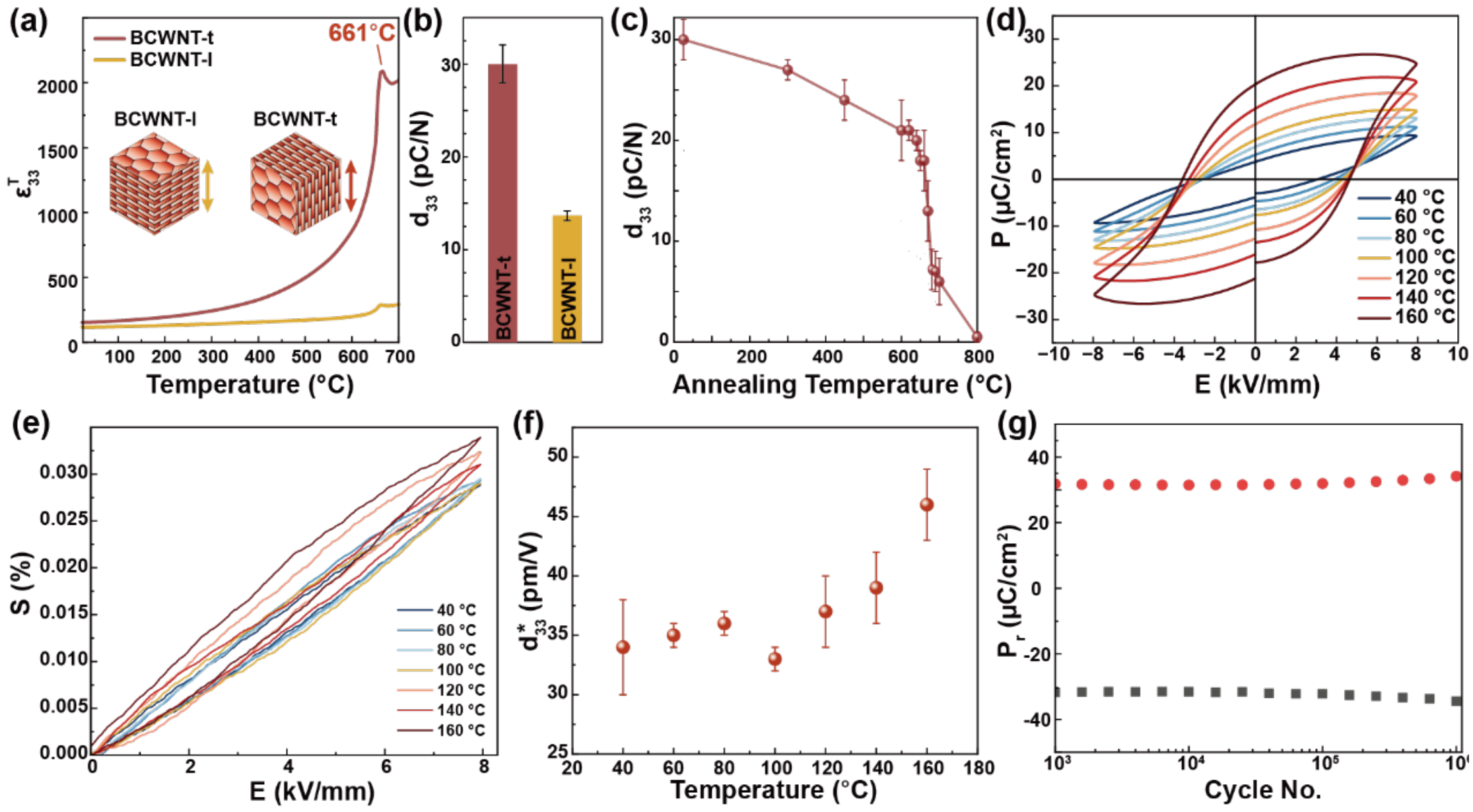

**Figure 3. Ferroelectric and piezoelectric performance improvement.** (a) Temperature-dependent permittivity pattern for all BCWNT samples. (b) $d_{33}$ value of BCWNT-t and -l at room temperature. (c) Temperature dependant *ex situ* $d_{33}$ value of BCWNT−$\boldsymbol{t}$ poled under 8 kV/mm, the inserted chart shows the $d_{33}$ value collections for BCWNT−$\boldsymbol{l}$ / −$\boldsymbol{t}$ poled under 8 kV/mm at 120 °C temperature, respectively. The samples were measured after annealing for 30 minutes at the given temperature (d) Temperature-dependant bipolar P-E loop for the BCWNT−$\boldsymbol{t}$ under 10 Hz. (e) Unipolar S-E loop at room temperature for BCWNT−$\boldsymbol{t}$, the sample is poled at 120 °C, 8 kV/mm. (f) Temperature-dependant $d_{33}^*$ calculated based on the unipolar S-E loop. (g) Bipolar P-E loop ferroelectric fatigue performance conducted at 10 Hz to record the loop at regular interval in between cycling 1k Hz, 180 °C, 7.5 kV/mm for BCWNT−$\boldsymbol{t}$.

Based on the above analysis, the material exhibited high fracture toughness in the -$\boldsymbol{tl}$ direction, and an optimum piezoelectric response is obtained in -$\boldsymbol{t}$ direction. This design allows piezoelectric output to be generated in one direction while the top and bottom surfaces primarily resist crack propagation from stress.

The enhancement in toughness and piezoelectric performance is predominantly achieved through the alignment of platelets through processing. Consequently, it can be extended to a wide range of BiT-based systems and, more generally, to other material derived from platelet powders. To verify this, we introduced the same nacre-like architecture into pure BiT ($<$ $00l>$ -textured NL-BiT, density > 97% Figure S) and observed similar substantial improvements. For toughness, the NL-BiT exhibits both intrinsic and extrinsic toughening with fracture plane parallel to texturing direction. Its $K_{IC}$ reaches 2.4 ± 0.3 MPa*m$^{0.5}$, while $K_J$ reaches 4.1 ± 0.1 MPa×m$^{0.5}$ (Figure S). These enhancements are comparable to those observed in NL-BCWNT. In addition, the textured NL-BiT shows $d_{33}$ of 13 ± 1 pC/N at transverse direction of textured ceramic (Figure S), consistent with previous reports[10,36]. The pronounced crack deflection is also observed in Figure Sb, c, indicating the crack deflection brings the toughening effect. Its $K_{IC}$ is slightly higher than that of NL-BCWNT (2.5 ± 0.1 MPa×m$^{0.5}$ vs 2.2 ± 0.4 MPa×m$^{0.5}$). This may indicate that doping reduces the intrinsic toughness of BCWNT compared with BiT, consistent with trend reported in the literature[12].

Figure 4a compares the toughness of our samples with those of other representative piezoelectric materials. Most of them exhibit relatively low fracture toughness, typically in the range of 0.8-2 MPa*m$^{0.5}$, indicating a limited resistance to crack initiation and propagation during electro-mechanical use. Once the nacre-like architecture is introduced, the toughness increases significantly to 4-5 MPa*m$^{0.5}$. This substantial improvement arises from stable crack-deflection toughening, which enables BiT-based materials to more effectively resist crack growth. To the best of our knowledge, this is likely to be the highest fracture toughness ever reported for piezoelectric ceramics.

Furthermore, the overall functional performance remains excellent, as shown in Figure 4b. Both $d_{33}$ and the Curie temperature remain high, outperforming some doped systems. Our approach therefore achieves simultaneous enhancements of mechanical properties without any sacrifice on the piezoelectric response. However, for the same Ce / W / Nb-doped BiT composition[7], the NL-BCWNT exhibited a slightly lower $d_{33}$ compared to the conventional ceramic. This reduction may originate from the lower resistivity and higher leakage of the MSS-derived platelets used in our work. Kimura et al.[45] suggested that residual $Na^+$, $K^+$ and $Cl^-$ ions, as well as adsorbed ionic species, can remain on the surface of powders after molten-salt synthesis. These species may promote abnormal conduction, increase dielectric loss and reduce the resistivity of the sintered ceramics.

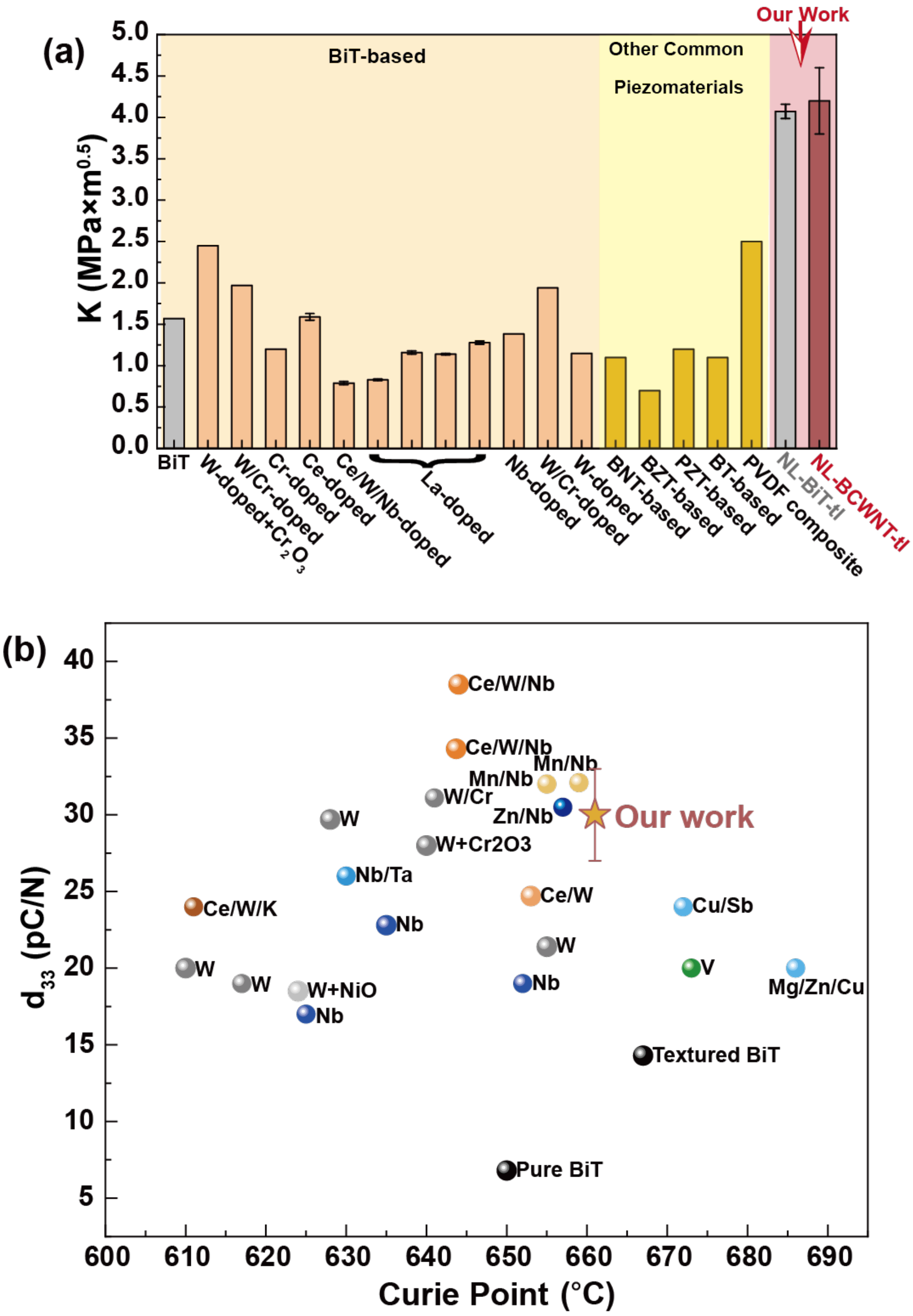

**Figure 4. Performance comparisons with existing piezoelectric materials.** (a) The fracture toughness comparisons with BiT-based ceramics[12,46–51] and other popular piezoelectric materials[3,52,53]. (b) The $d_{33}$ and Curie points values collected for BiT-based high temperature piezoceramics[7,10,36,40,54–60].

### *Conclusion*

In this work, we construct a $<00l>$ –fibre textured (Lotgering factor $f_{001}$ > 94%) nacre-like BCWNT ceramic. This structural design decouples the direction of best mechanical and piezoelectric properties, thereby providing directional performance requirements encountered in some real-world devices. This structure gives BCWNT ceramics enhanced mechanical performance. The flexural strength when the sample is oriented so that the tensile stress is applied in the transverse direction and the direction of maximum opening stress is applied along the longitudinal and transverse direction are 80 ± 3 MPa and 90 ± 20 MPa, respectively. A high toughness compared to literature values of the same composition is obtained with $K_{IC}$ 2.2 ± 0.4 MPa·m$^{0.5}$ and $K_J$ of 4.2 ± 0.4 MPa·m$^{0.5}$. The increase in toughness at initiation and presence of stable crack propagation is obtained through crack deflection from the nacre-like structure along the texture directions. Good piezoelectric performances are also obtained in the transverse direction. The $d_{33}$ reaches 30 ± 3 pC/N and $d_{33}^*$ is 46 ± 3 pm/V with stable value with low remnant strain in unipolar SE loop. These enhancements come from the multiple dopants that influence domain size and mobility along the aligned spontaneous polarization direction $< h00 >$ within the textured ceramics. The Curie point was found to be 661°C. NL-BCWNT maintained an excellent lifetime and fatigue performance higher than $10^6$ bipolar P-E loops without any measurable degradation. Moreover, same improvements were also obtained in pure BiT system, indicating that this bioinspired strategy could be applicable to other piezoelectric material systems for applications where both functional and structural performance are critical, enriching the toolbox for electromechanical materials designs.

## Materials and methods

### $Bi_{3.96}Ce_{0.04}Ti_{2.965}W_{0.0175}Nb_{0.0175}O_{12}$ platelets powder Molten Salt Synthesis (MSS)

The first step is synthesizing the Aurivillius phase BCWNT with a solid state reaction: The $Bi_2O_3$ (Sigma-Aldrich, 99.9%), $TiO_2$ (Sigma-Aldrich, 99.8%), $CeO_2$ (Sigma-Aldrich, ⩾99%), $Nb_2O_5$ (Sigma-Aldrich, 99.9%) and $WO_3$ (Sigma-Aldrich, 99.9%) were mixed according to the stoichiometric composition, except $Bi_2O_3$ for which a 3 wt% excess was added to compensate for its evaporation during heat treatment. The mixture was ball milled with zirconia milling balls in ethanol (same weight as mixture) for 24 - 48 h, dried and calcinated under 850 for 4 h. Then part of the calcinated powder was mixed with a molten salt mixture of NaCl (VWR, >99.5%) and KCl (Sigma-Aldrich, > 99.0%) with a molar ratio 1:1. The calcined oxides and salt mixture were mixed with a weight ratio of 1:1 with ethanol and ball milled for 24 h with zirconia milling balls. The well broken and mixed mixtures were dried and then heated at 1030 °C with a 2 h dwell and heating rate of 5 °C/min. The obtained mixture was washed by deionized water and dried under 120 °C for 24 hours.

### BCWNT ceramic shaping with MASC and sintering

The platelets were coated with superparamagnetic iron oxide nanoparticles (SPIONS, ferrofluid EMG605 from FerroTec): the powders were first dispersed in water at 50 ml/g and stirred to obtain a uniform suspension; EMG605 was then added at 10 μl/g, followed by stirring for 24 h. The powders were collected by vacuum filtration, separated from the aqueous medium, and dried in an oven at 120 °C for 3 h. Next, for the magnetically assisted slip casting, an ethanol-based slurry containing $Fe_3O_4$-coated BCWNT platelets at a solid of 15 vol% was prepared. Polyvinylpyrrolidone was added as a binder at 5 wt% relative to the mass of BCWNT. The slurry was magnetically stirred for 24 h. The slurry was poured into a cylindrical mold (Ø3 cm × 5 cm) which rest on top of a Polydimethylsiloxane flat substrate. Casting gypsum ("Plaster of Paris") was used to make the plate and prepared according to the provider instruction. A 500 mT permanent neodymium magnet (NIBL 00954, N38H grade, Magnet Sales) was rotated horizontally at a rotation speed > 250 rpm to align the platelets horizontally within this slurry and positioned around 2 cm away from the mold. After dried, the green body was transferred to a box furnace, debinded at 550 °C for 4 h, then moved into SPS furnace (KCE®-FCT HP D 10-SD from FCT System) in a graphite mold (Ø3 cm), and was sintered under argon at 1030 °C with a dwell of 5 min and a heating rate of 50 °C /min. The sintered NL-BCWNT were then polished to remove the graphite paper and annealed at 800 - 820 °C in air for 3 h to remove carbon contaminations.

### Structural characterization

The phases and chemical compounds were measured by X-ray diffraction (XRD) via MPB Panalytical. The Lotgering Factor[4] was used to evaluate the $\langle 00l \rangle$ texturing quality, which was calculated based on XRD according to following equation:

$$F_{(00l)} = \frac{P - P_0}{1 - P_0} \quad (1)$$

$$P = \frac{\sum I_{(00l)}}{\sum I_{(hkl)}} \quad (2)$$

$$P_0 = \frac{\sum I_{0(00l)}}{\sum I_{0(hkl)}} \quad (3)$$

The $P_0$ and $P$ mean the intensity sum of target equivalent crystallographic direction families for random sample and textured samples respectively. The scanning electron microscope (SEM) figures, energy-dispersive X-ray spectroscopy (EDX) mapping and electron backscatter diffraction (EBSD) pattern were detected by TFS Quanta ESEM and Zeiss Sigma300. The grain orientations and microstructure were analyzed by EBSD pole figure and inverse pole figure reindexed by MTEX. EBSD data and pole figures were plotted based on these data. The powder size distribution and grain size distribution were quantitively analyzed by SEM and ImageJ.

**Mechanical property characterization.**

The strength and toughness were determined by in-situ three-point bending, with a span of S = 20 mm. For toughness, single-edge-notched beam (SENB) toughness test specimens were cut from the sintered pellets. The ceramics were machined into beams with a length of 30 - 20 mm, thickness of W = 3 - 5 mm and width of B = 3 - 4 mm (B<W). The notch length was between 0.4W < a < 0.6W, with width of 0.5 mm, and the notch tip was sharpened according to ASTM E1820. The three-point bending fracture toughness was tested in a 10kN universal testing machine (Z010, ZwickRoell) under constant displacement rate of 1 μm/s. For the calculation, the critical stress intensity factor $K_{IC}$ was calculated using ASTM E1820:

$$K_{IC} = \left[\frac{FS}{BW^{3/2}}\right] \times f\left(\frac{a}{W}\right) \quad (4)$$

where F is the applied force when the fracture occurs, and $f$ is given by:

$$f\left(\frac{a}{W}\right) = \frac{3(\frac{a}{W})^{\frac{1}{2}} \times [1.99 - \frac{a}{W}(1-\frac{a}{W})(2.15 - 3.93\frac{a}{W} + 2.7\left(\frac{a}{W}\right)^2)]}{2(1+2\frac{a}{W})(1-\frac{a}{W})^{3/2}} \quad (5)$$

To determine crack resistance curves, nonlinear elastic fracture mechanics was considered, including the contribution of elastic and plastic deformation. This was calculated according to ASTM E1820 with J-integral method, where the elastic contribution $J_{el}$ was given by:

$$J_{el} = \frac{K_i^2}{E'} \quad (6)$$

where $K_i$ is the fracture toughness at a given crack length $a_i$ and its corresponding force $F_i$, $E'$ is $E/(1-\upsilon^2)$, $E$ is the Young's modulus and $\upsilon$ is Poisson's ratio (0.25 here based on previous reports[61]). The $J_{pl}$ is calculated iteratively using the following expression:

$$J_{pl(i)} = [J_{pl(i-1)} + \frac{1.9\left(A_{pl(i)} - A_{pl(i-1)}\right)}{b_{i-1}B}] \times \left[1 - \frac{0.9 \times (a_i - a_{i-1})}{b_{i-1}}\right] (7)$$

where $b_{i-1}$ is the uncracked ligament width. The $A_{pl}$ was evaluated at each increment of crack length, with the plastic area given by:

$$A_{pl(i)} = A_{pl(i-1)} + (F_i + F_{i-1})\frac{\upsilon_{pl(i)} - \upsilon_{pl(i-1)}}{2} \quad (8)$$

where $\upsilon_{pl(i)}$ is the plastic part of the force-displacement given by $\upsilon_{(i)} - F_i C_i$, $\upsilon_{(i)}$ is the displacement and the compliance $C_i$ was determined from:

$$C_i = \frac{1}{EB}(\frac{S}{W-a_i})^2 \times \left[1.193 - 1.98\frac{a_i}{W} + 4.478\left(\frac{a_i}{W}\right)^2 - 4.443\left(\frac{a_i}{W}\right)^3 + 1.739\left(\frac{a_i}{W}\right)^4\right] (9)$$

The J integral values are corrected using the following expression:

$$J = J_{el} + \frac{J_{pl}}{1 + (\frac{\alpha - 0.5}{\alpha + 0.5})\frac{\Delta a}{b}} (10)$$

where $\alpha$ is 1 for SENB specimen, and b is uncracked ligament width. The stress intensity $K_J$ is calculated using the following expression:

$$K_J = \sqrt{JE'} (11)$$

To measure the strength, the ceramics were machined into beams with a length of 30 - 20 mm, thickness of W = 3 - 5 mm and width of B = 4 - 6 mm (B>W), the bottom surfaces were polished into ~1 µm and the edges were chamfered. Three-point fracture bending tests assisted by *in situ* optical microscopy imaging were carried out to measure the sample deflection. The optical setup consists of a 31-megapixel camera with a high speed 10 Gbit/s interface (Ory 10GigE, Teledyne FLIR) and a high resolution telecentric lens (resolution ~5 µm/px, VS-LTC3.3-45/FS, VS Technology). The beams were illuminated with a low angle diffused ring light (VL-LRD73100W, VS Technology), and the images were collected at 1.98Hz. The sample bending displacements 3D drifting corrected by ImageJ[62]. The strength $\sigma_f$ and strain $\delta$ were calculated following:

$$\sigma_f = \frac{3FL}{2BW^2} (12)$$

$$\varepsilon = \frac{6\delta W}{L^2} (13)$$

The Young's modules were determined by the slope of the stress-strain curve. The hardness was determined by Vickers hardness (Indentec, ZwickRoell) under 5 kg and holding for 10 s on well-polished samples (~1 µm).

### Piezo-, ferro- and dielectric testing

The samples for functional characterizations were machined out from the sintered disks, made into round disks with diameters of 3-5 mm and thickness of 0.4 - 0.5 mm. Electrodes were deposited onto the upper and lower end faces of the disk-shaped specimens as follows. First, a layer of conductive silver paste (FuelCellMaterials, FCM 233006 Ag-I) was applied to both end faces and preliminarily dried at 120 °C for 20 minutes. Thereafter, an additional coat of silver paste was applied, and the samples were fired at 650 °C for 30 minutes to ensure robust adhesion of the electrodes.

To measure the piezoelectric properties of the ceramics, samples were poled in a silicone oil bath by applying a DC electric field of 5 kV/mm (BCWNT-reference sample) and 8 kV/mm (NL-BCWNT) at 120 °C for 30 minutes. A ferroelectric testing system with a laser interferometer (TF1000, aixACCT Systems GmbH, Germany) was used to evaluate the field-induced electromechanical (strain-field and polarisation-field) response from room temperature to 160 °C. The piezoelectric coefficient $d_{33}$ was measured using a $d_{33}$ meter (APC International). The *ex situ* $d_{33}$ evolution with temperature was performed by annealing the samples at various temperatures up to 700 °C, with a hold time of 30 minutes, followed by measurements of the $d_{33}$ at room temperature. 3-5 samples were used for each characterization. The bipolar fatigue testing was done on the TF1000 with a triangular wave form with amplitude of 4 and 7.5 kV/mm at 180 °C and a frequency of 1000 Hz, with intermittent P-E loops recorded at 10 Hz.

## Acknowledgment

The authors would like to thank Dr James Roscow and Dr Hamideh Khanbareh from the University of Bath for their advices on the piezoelectric poling and $d_{33}$ measurements in this work.

SUPPORT INFORMATION

**Table S1** Particle morphology statistic value for BCWNT and BiT platelets.

| Samples | Median diameter $D_{50}$ (μm) | Median thickness $t_{50}$ (μm) | Median aspect ratio $A_{50}$ (-) |
|---|---|---|---|
| BCWNT | 3.4 | 0.2 | 15 |
| BiT | 4.2 | 0.3 | 12 |

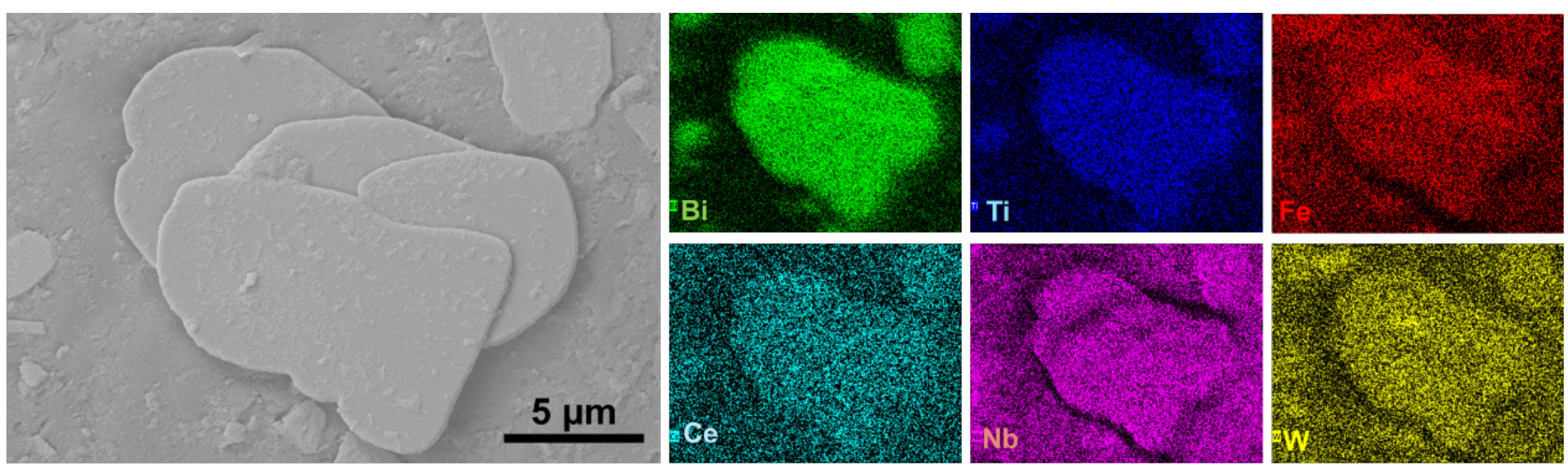

**Figure S1** Energy-dispersive X-ray spectroscopy (EDX) mapping for $Fe_3O_4$-coated BCWNT platelets powder.

**Table S2** Density and relative density for BCWNT ceramic samples.

| Samples | Sintering | Density (g/cm$^3$) | Relative Density (%) |
|---|---|---|---|
| NL-BCWNT | SPS | 7.3 ± 0.9 | 91.0 ± 0.7 |

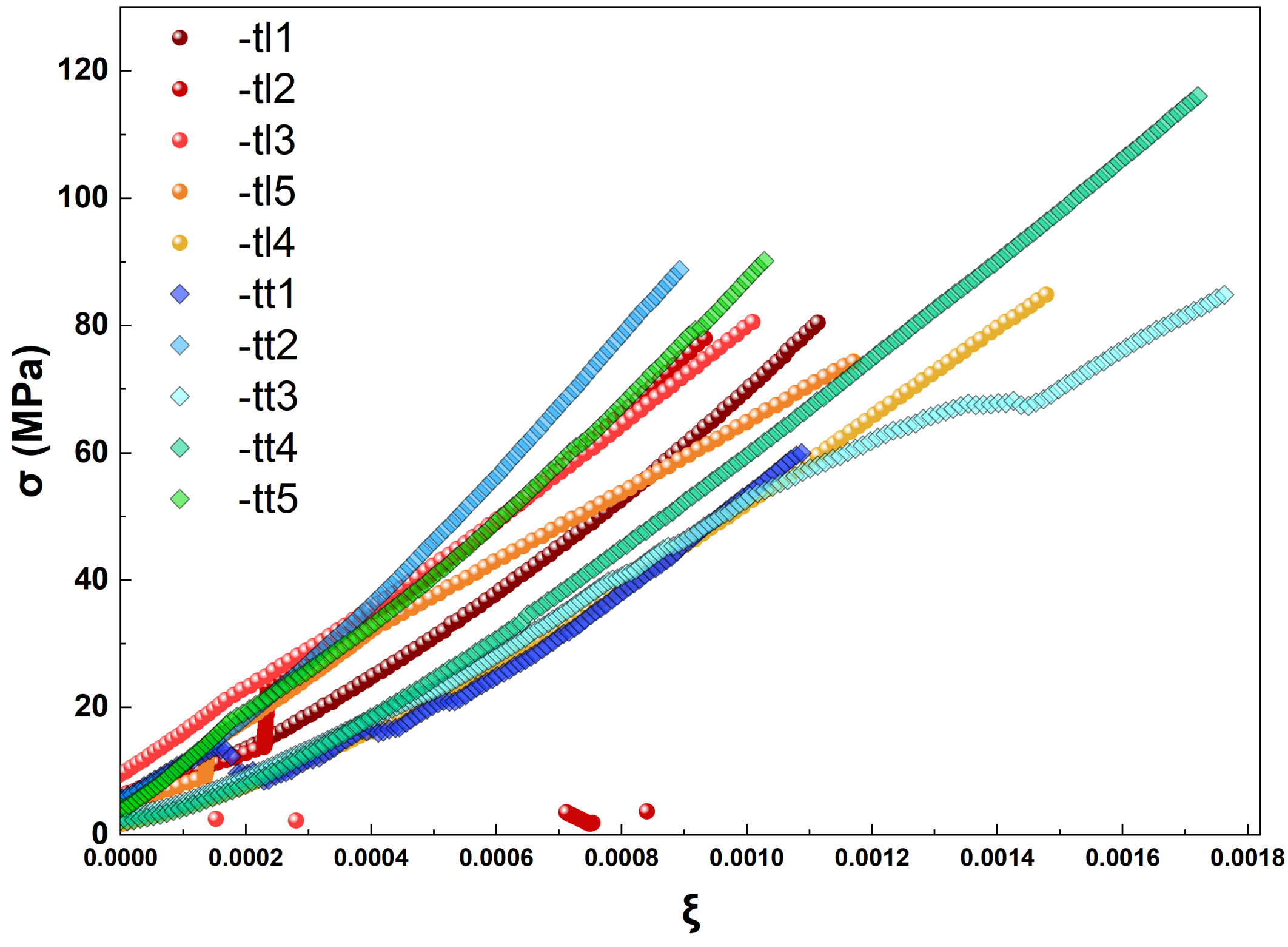

Figure S2. Stress-strain curve for BCWNT ceramics. All the strains were calibrated.

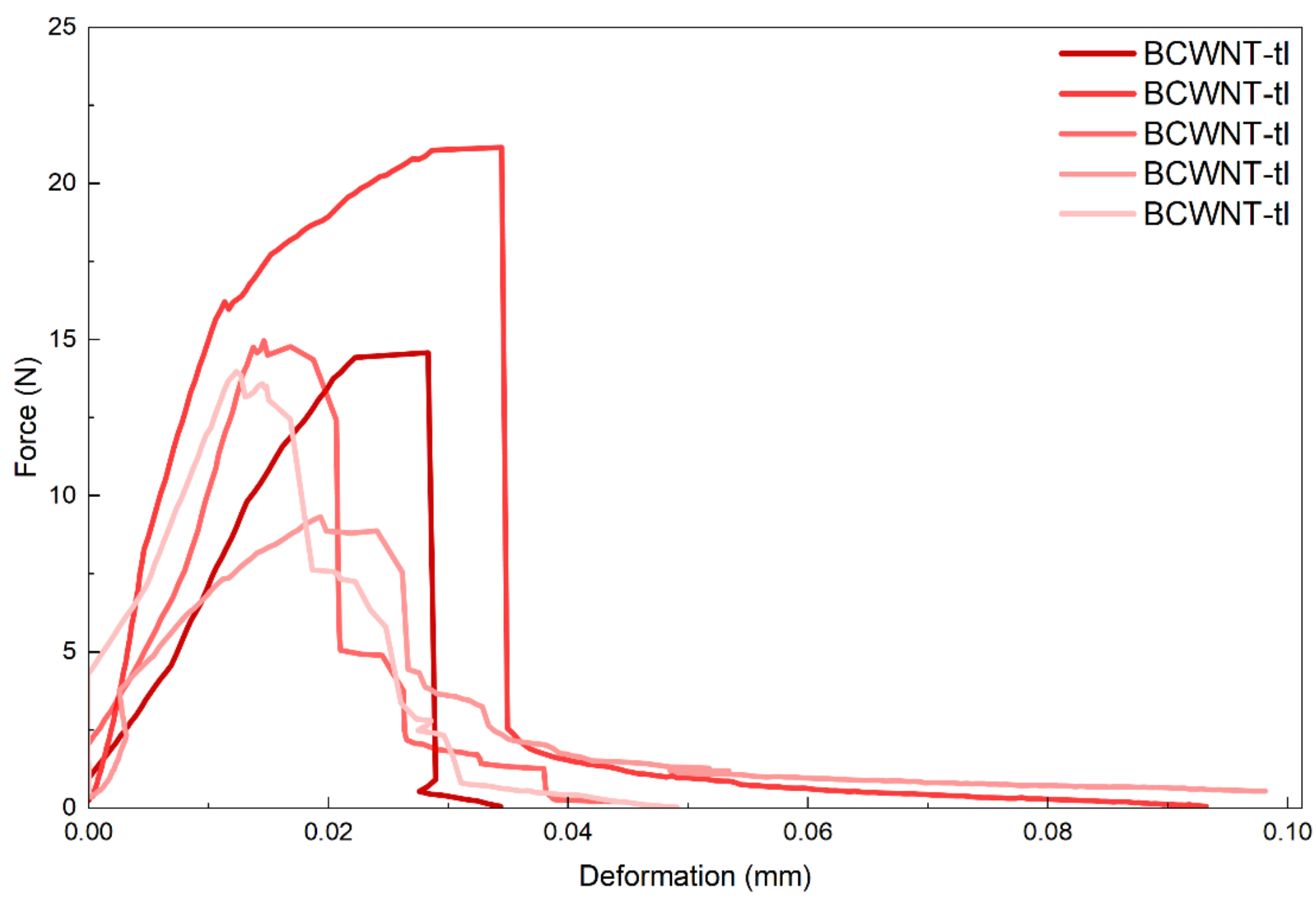

**Figure S3. Force-displacement of SENB fracture toughness test for BCWNT ceramics.** The displacements have been calibrated.

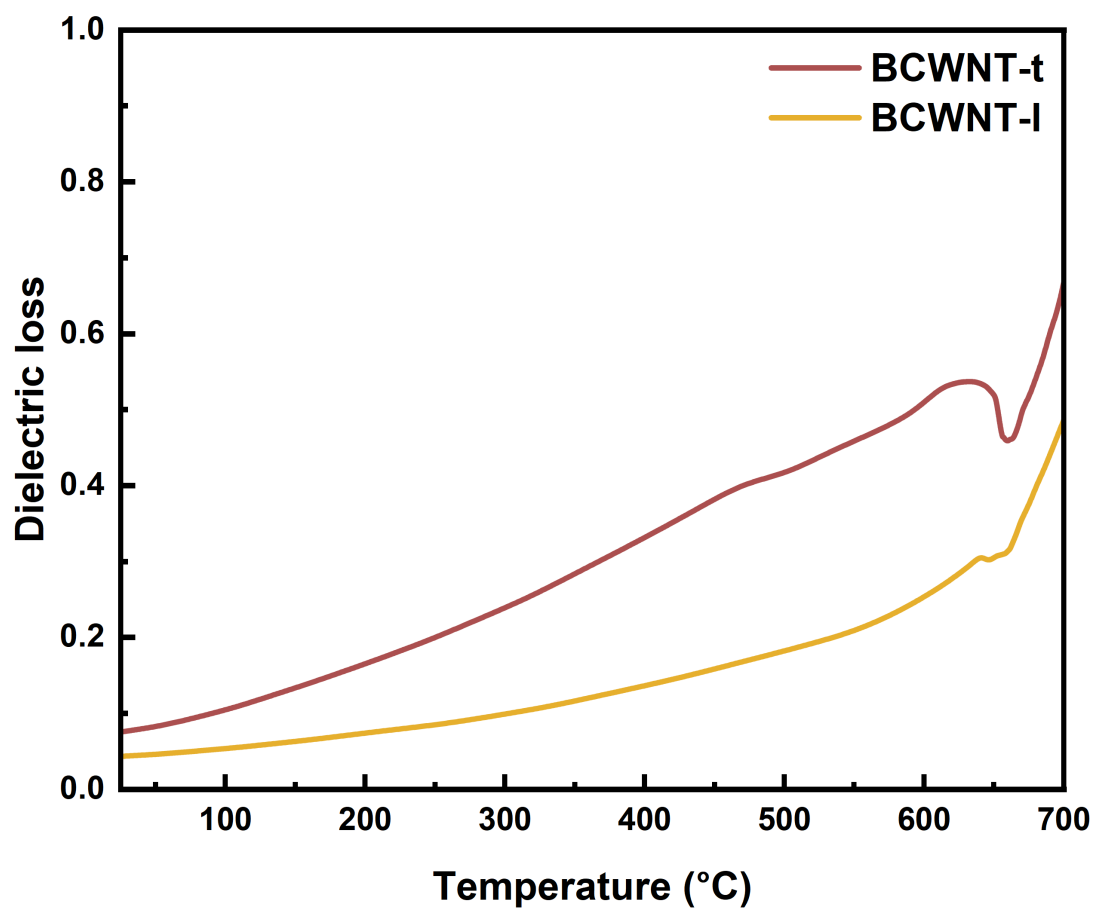

**Figure S4** Dielectric loss for all BCWNT samples.

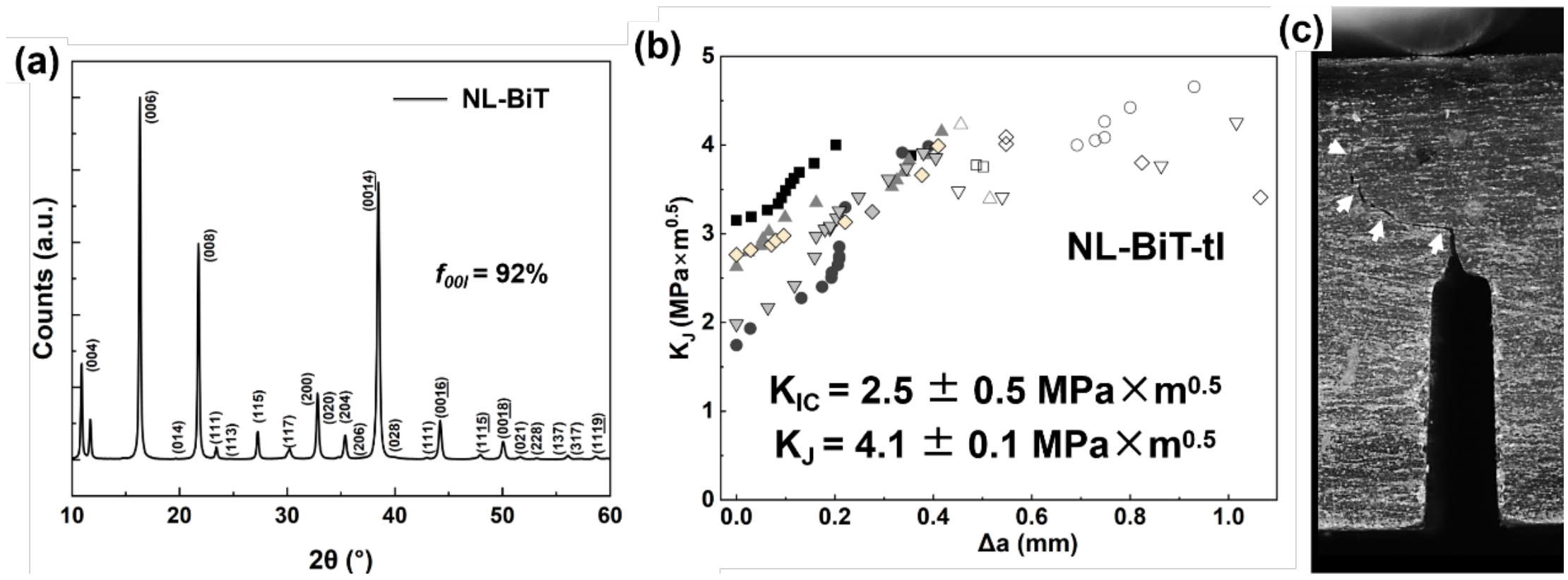

**Figure S5 SENB fracture-toughness measurement of NL-BiT-tl.** (a) XRD pattern of textured NL-BiT, showing enhanced <00l>-peaks intensities, indicating the <00l>-textured orientation. (b) R-curves for different samples, showing the calculated $K_{IC}$ and $K_J$ values. (c) Crack-growth path corresponding to the NL-BiT-tl test.

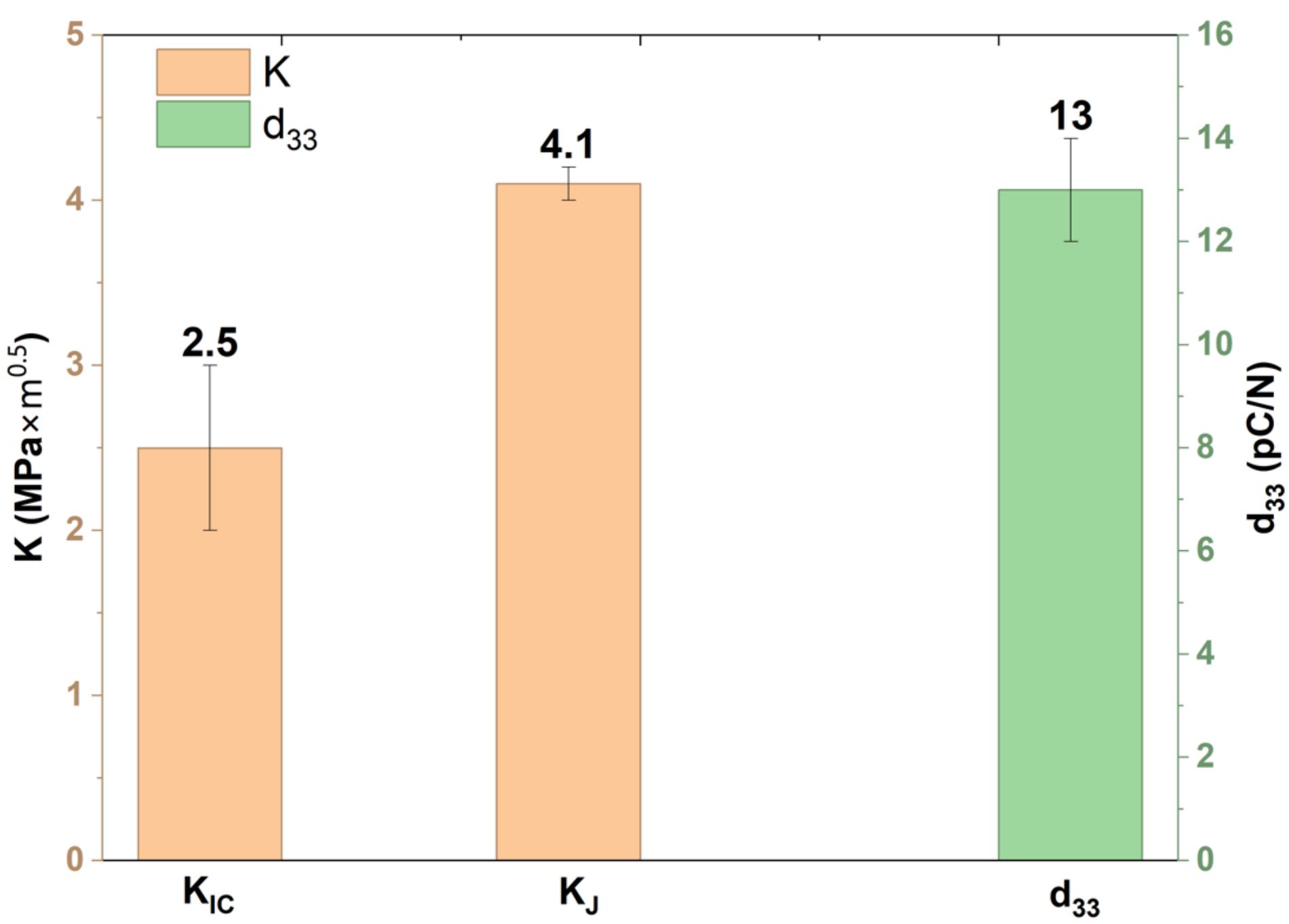

**Figure S6** Toughness and piezoelectric performance for undoped NL-BiT at -tl direction and -t direction, respectively.